\documentclass[letterpaper,twocolumn,letter]{jpsj3}
\usepackage{txfonts}
\usepackage[dvipdfmx]{graphicx}
\usepackage{color}
\usepackage{ulem}

\usepackage[final]{changes}

\usepackage{txfonts}
\usepackage{bm}

\title{First Observation of de Haas-van Alphen Effect and Fermi Surfaces in Unconventional Superconductor UTe$_2$}

\author{
Dai~Aoki$^1$\thanks{E-mail: aoki@imr.tohoku.ac.jp}, 
Hironori~Sakai$^2$,
Petr~Opletal$^2$,
Yoshifumi~Tokiwa$^2$,
Jun~Ishizuka$^3$,
Youichi~Yanase$^4$,
Hisatomo~Harima$^5$,
Ai~Nakamura$^1$,
Dexin~Li$^1$,
Yoshiya~Homma$^1$
Yusei~Shimizu$^1$,
Georg~Knebel$^6$,
Jacques~Flouquet$^6$,
and Yoshinori~Haga$^2$
}

\inst{%
$^1$IMR, Tohoku University, Oarai, Ibaraki 311-1313, Japan\\
$^2$Advanced Science Research Center, Japan Atomic Energy Agency, Tokai, Ibaraki 319-1195, Japan\\
$^3$Faculty of Engineering, Niigata University, Ikarashi, Niigata 950-2181, Japan\\
$^4$Department of Physics, Graduate School of Science, Kyoto University, Kyoto 606-8502, Japan\\
$^5$Graduate School of Science, Kobe University, Kobe 657-8501, Japan\\
$^6$Univ. Grenoble Alpes, CEA, Grenoble INP, IRIG, PHELIQS, F-38000 Grenoble, France\\
}

\abst{%
We report the first observation of the de Haas-van Alphen (dHvA) effect in the novel spin-triplet superconductor UTe$_2$ using high quality single crystals with the high residual resistivity ratio (RRR) over 200. 
The dHvA frequencies, named $\alpha$ and $\beta$, are detected for the field directions between $c$ and $a$-axes. 
The frequency of branch $\beta$ increases rapidly with the field angle tilted from $c$ to $a$-axis,
while branch $\alpha$ splits, owing to the maximal and minimal cross-sectional areas from the same Fermi surface.
Both dHvA branches, $\alpha$ and $\beta$ reveal two kinds of cylindrical Fermi surfaces with a strong corrugation at least for branch $\alpha$.
The angular dependence of the dHvA frequencies is in very good agreement with that calculated by 
the generalized gradient approximation (GGA) method taking into account the on-site Coulomb repulsion of $U=2\,{\rm eV}$,
indicating the main Fermi surfaces are experimentally detected.
The detected cyclotron effective masses are large in the range from $32$ to $57\,m_0$.
They are approximately 10--20 times lager than the corresponding band masses,
consistent with the mass enhancement obtained from the Sommerfeld coefficient, $\gamma$ and the calculated density of states at the Fermi level.
The local density approximation (LDA) calculations of ThTe$_2$ assuming U$^{4+}$ with the 5$f^2$ localized model are in less agreement with our experimental results, in spite of the prediction for two cylindrical Fermi surfaces,
suggesting a mixed valence states of U$^{4+}$ and U$^{3+}$ in UTe$_2$.
}

\begin{document}
\maketitle
UTe$_2$ attracts much attention because of the unusual superconducting properties due to the spin-triplet state.~\cite{Ran19,Aok19_UTe2,Aok22_UTe2_review}
UTe$_2$ is a heavy fermion paramagnet with a Sommerfeld coefficient $\gamma\sim 120\,{\rm mJ K^{-2}mol^{-1}}$.
It crystallizes in the body-centered orthorhombic structure with the space group, $Immm$ (No. 71, $D_{2h}^{25}$),
where the U atom forms two-leg ladder structure along the $a$-axis.
Superconductivity occurs below $T_{\rm c}=1.5$--$2\,{\rm K}$ with the large specific heat jump.~\cite{Ros22} 
A highlight is the huge upper critical field $H_{\rm c2}$ with a field-reentrant behavior for $H\parallel b$-axis (hard-magnetization axis). Superconductivity survives up to the first-order metamagnetic transition at $H_{\rm m}=35\,{\rm T}$,~\cite{Kne19} as bulk properties~\cite{Rosuel22}.
The values of $H_{\rm c2}$ highly exceed the Pauli limit, ($\sim 3\,{\rm T}$), for all field directions, suggesting a spin-triplet state. 
A microscopic evidence for a spin-triplet state is obtained from NMR experiments, in which the Knight shift is unchanged or decreases very slightly below $T_{\rm c}$ for $H \parallel a$, $b$ and $c$-axis.~\cite{Nak21}
Another highlight is the appearance of the multiple superconducting phases under pressure.~\cite{Bra19,Aok20_UTe2}
Applying pressure, $T_{\rm c}$ starts decreasing, and splits at $\sim 0.3\,{\rm GPa}$. 
The lower $T_{\rm c}$ decreases continuously, while the higher $T_{\rm c}$ increases up to $3\,{\rm K}$ around $1\,{\rm GPa}$ then decreases rapidly. 
At the critical pressure $P_{\rm c}\sim 1.5\,{\rm GPa}$, superconductivity is suppressed and the magnetic order, most likely aniferromagnetic order appears above $P_{\rm c}$.~\cite{Bra19,Tho20,Aok20_UTe2,Kna21}
Under magnetic field,  multiple superconducting phases show the remarkable field response, displaying a sudden increase of $H_{\rm c2}$ at low temperatures.~\cite{Kne20,Aok20_UTe2}
These multiple superconducting phases are the hallmarks of a spin-triplet state with different superconducting order parameters related to the spin and orbital degree of freedom.

It was first pointed out~\cite{Ran19} that UTe$_2$ is located at the verge of the ferromagnetic order and resembles ferromagnetic superconductors~\cite{Aok12_JPSJ_review,Aok19}.
However, no ferromagnetic fluctuations are experimentally established; instead antiferromagnetic fluctuations with an incommensurate wave vector are detected in inelastic neutron scattering experiments.~\cite{Dua20,Kna21_PRB}.
Furthermore, above $P_{\rm c}$, antiferromagnetic order is confirmed directly by magnetic susceptibility measurements~\cite{Li21}.
These results suggest that multiple fluctuations, such as ferromagnetic, antiferromagnetic fluctuations, and valence instabilities~\cite{Tho20,Miy22}, exist in UTe$_2$, playing important roles for the unusual superconducting properties.

The electronic structure of UTe$_2$ has been investigated by angle-resolved photoemission spectroscopy (ARPES) at $20\,{\rm K}$.
The results obtained from the soft X-ray~\cite{Fuj19} and the vacuum ultraviolet synchrotron radiation~\cite{Mia20} are contradicting, probably related to the different inelastic mean free path of the photoelectrons.
The fine structure near the Fermi level was unresolved in the soft X-ray experiments.
On the other hand, the high resolution ARPES experiments revealed two light quasi-one dimensional bands and one heavy band at the Fermi level,
which is, however, inconsistent with the soft X-ray ARPES experiments.
Thus, no clear conclusion on the electronic structure emerges up to now.
The determination of the Fermi surface topology at low temperatures through the direct observation of quantum oscillations is highly desired,
which will be a key experiment to investigate the topological superconducting phenomena as well.

In order to clarify the electronic structure, we performed de Haas-van Alphen (dHvA) experiments on new high quality single crystals. 
Clear dHvA oscillations were successfully detected, and the angular dependence of the dHvA frequencies reveals
two kinds of cylindrical Fermi surfaces.
The results are well explained by the generalized gradient approximation (GGA) with the on-site
Coulomb repulsion, $U=2\,{\rm eV}$.
The detected large cyclotron effective masses are consistent with the Sommerfeld coefficient.

High quality single crystals of UTe$_2$ were grown at Oarai (sample \#1) and Tokai (sample \#2).
The details of single crystal growth will be published elsewhere~\cite{Sak22}.
The dHvA experiments were performed at low temperatures down to $70\,{\rm mK}$ and at high fields up to $147\,{\rm kOe}$, as well as resistivity, specific heat and AC susceptibility measurements.~\cite{sup1}
The band calculations were done by the GGA+$U$ method in UTe$_2$~\cite{Ish19,Ish19a} and the local density approximation (LDA) method in ThTe$_2$ as a reference.~\cite{Har20}

\begin{figure}[tbh]
\begin{center}
\includegraphics[width= 0.8\hsize,clip]{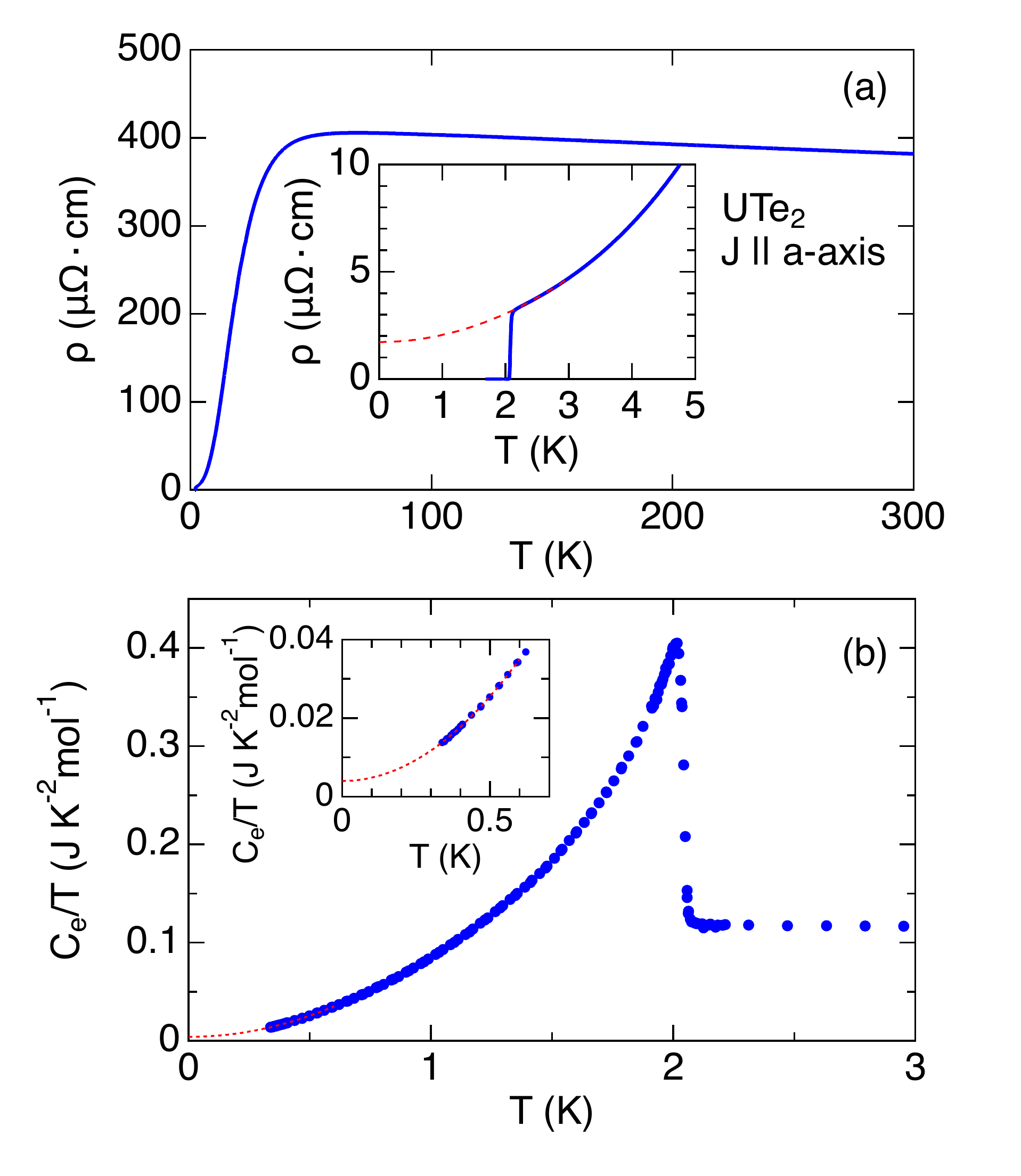}
\end{center}
\caption{(Color online) (a) Temperature dependence of the resistivity for the current along $a$-axis in UTe$_2$ (sample \#1). The residual resistivity ratio (RRR) is 220. The dotted line is the results of fitting.
(b) Temperature dependence of the electronic specific heat in the form of $C_{\rm e}/T$ vs $T$ in UTe$_2$ (sample \#1). The dotted line is the results of fitting between $0.34$ and $0.6\,{\rm K}$ assuming $C_{\rm e}/T = \gamma_0 + B T^2$.}
\label{fig:UTe2_resist_Cp}
\end{figure}

First we present the superconducting properties of our high quality single crystals. 
Figure~\ref{fig:UTe2_resist_Cp}(a) shows the temperature dependence of the resistivity for the current along the $a$-axis. 
The superconducting transition temperature, $T_{\rm c}=2.06\,{\rm K}$ defined by zero resistivity is very high and very sharp.
The resistivity follows the $T^2$ dependence at low temperatures below $3.5\,{\rm K}$.
The residual resistivity $\rho_0$ and the residual resistivity ratio RRR ($\equiv \rho_{300 \rm K}/\rho_0$) 
are $1.7\,\mu\Omega\!\cdot\!{\rm cm}$ and 220, respectively.

Figure~\ref{fig:UTe2_resist_Cp}(b) shows the temperature dependence of the electronic specific heat in the form of $C_{\rm e}/T$ vs $T$ for sample\#1 after subtracting the phonon contribution.
A sharp and large single-jump at $T_{\rm c}=2.05\,{\rm K}$ with the width of $0.04\,{\rm K}$ and $\Delta C_{\rm e}/(\gamma T_{\rm c}) = 2.64$ associated with a small residual $\gamma$-value ($\gamma_0$), which is only $3\,{\%}$ of the normal state $\gamma$-value ($\gamma_{\rm N}$), is observed.
The small $\gamma_0$ and high $T_{\rm c}$ in sample\#1 are compared to those in different quality samples.~\cite{sup2}
All these properties indicate the high quality of our dHvA samples.

\begin{figure}[tbh]
\begin{center}
\includegraphics[width= \hsize,clip]{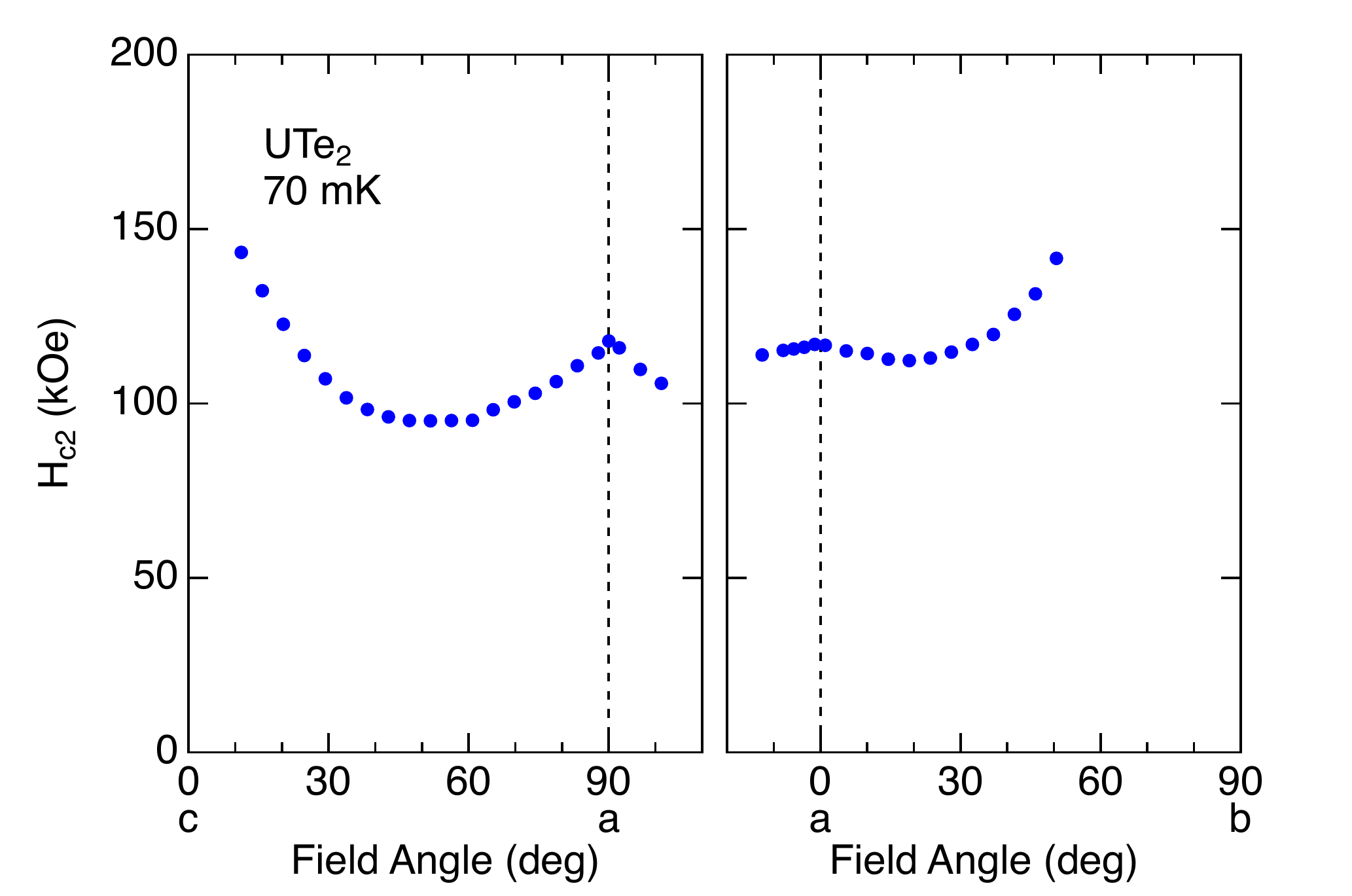}
\end{center}
\caption{(Color online) Anisotropy of $H_{\rm c2}$ for the field direction from $c$ to $a$-axis, and from $a$ to $b$-axis at $70\,{\rm mK}$ determined by the AC susceptibility measurements in UTe$_2$ (sample\#2)}
\label{fig:UTe2_Hc2_AngDep}
\end{figure}

Figure~\ref{fig:UTe2_Hc2_AngDep} shows the anisotropy of $H_{\rm c2}$ at $70\,{\rm mK}$ for the field directions from $c$ to $a$-axis, and from $a$ to $c$-axis using the dHvA sample (\#2) with $T_{\rm c}=2.05\,{\rm K}$.
Because of the high $T_{\rm c}$, $H_{\rm c2}$ shifts to the higher field, compared to the previous results~\cite{Aok19_UTe2}.
$H_{\rm c2}$ for $a$-axis reaches $118\,{\rm kOe}$, and $H_{\rm c2}$ for $c$-axis exceeds our highest field $147\,{\rm kOe}$, most likely around $160\,{\rm kOe}$.
These high $H_{\rm c2}$ values restrict our dHvA experiments, since the dHvA oscillations appears above $H_{\rm c2}$ as shown later. 
The unusual minima were found around $50\,{\rm deg}$ tilted from $c$ to $a$-axis, and $20\,{\rm deg}$ from $a$ to $b$-axis, associated with a sharp maximum for $H\parallel a$-axis.
This is probably a mark of the field variation of the pairing strength for $a$-axis, which should be clarified in the temperature dependence of $H_{\rm c2}$ with angular singularities through the precise experiments.

\begin{figure}[tbh]
\begin{center}
\includegraphics[width= 0.9\hsize,clip]{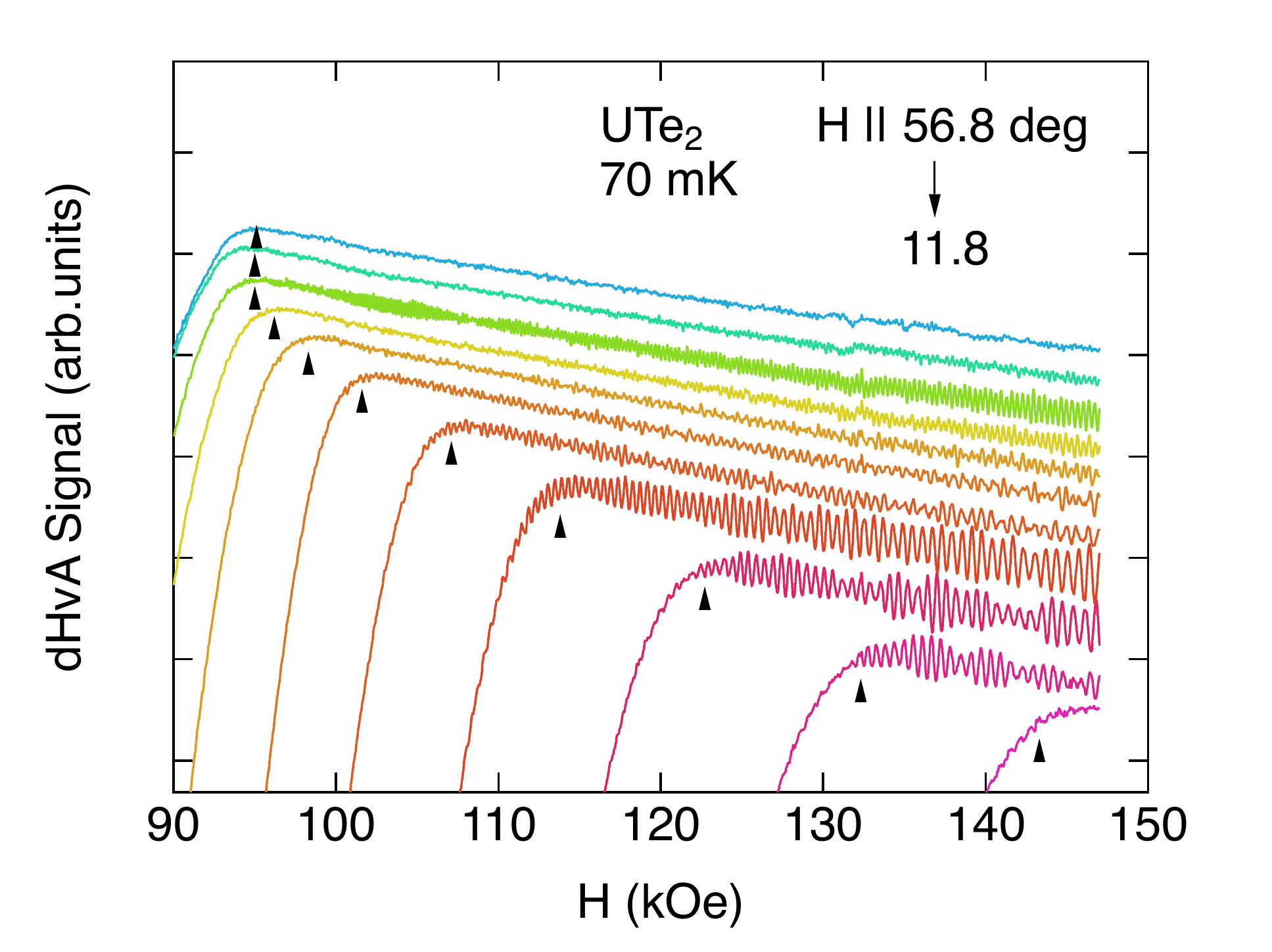}
\end{center}
\caption{(Color online) dHvA oscillations at $70\,{\rm mK}$ at different field angles with the $4.5\,{\rm deg}$ step tilted from $c$ to $a$-axis in UTe$_2$ (sample\#2). Small up-arrows indicate $H_{\rm c2}$.}
\label{fig:UTe2_osc_ang}
\end{figure}

Next we show in Fig.~\ref{fig:UTe2_osc_ang} the dHvA oscillations at different field angles tilted from $c$ to $a$-axis.
The clear dHvA signals were observed at the field angles between $11.8$ and $56.8\,{\rm deg}$ above $H_{\rm c2}$ denoted by up-arrows.
Even below $H_{\rm c2}$, the dHvA oscillations are observed, but the amplitudes are strongly damped.
At higher field angles close to $a$-axis, no dHvA oscillations were detected. 

\begin{figure}[tbh]
\begin{center}
\includegraphics[width= 0.8\hsize,clip]{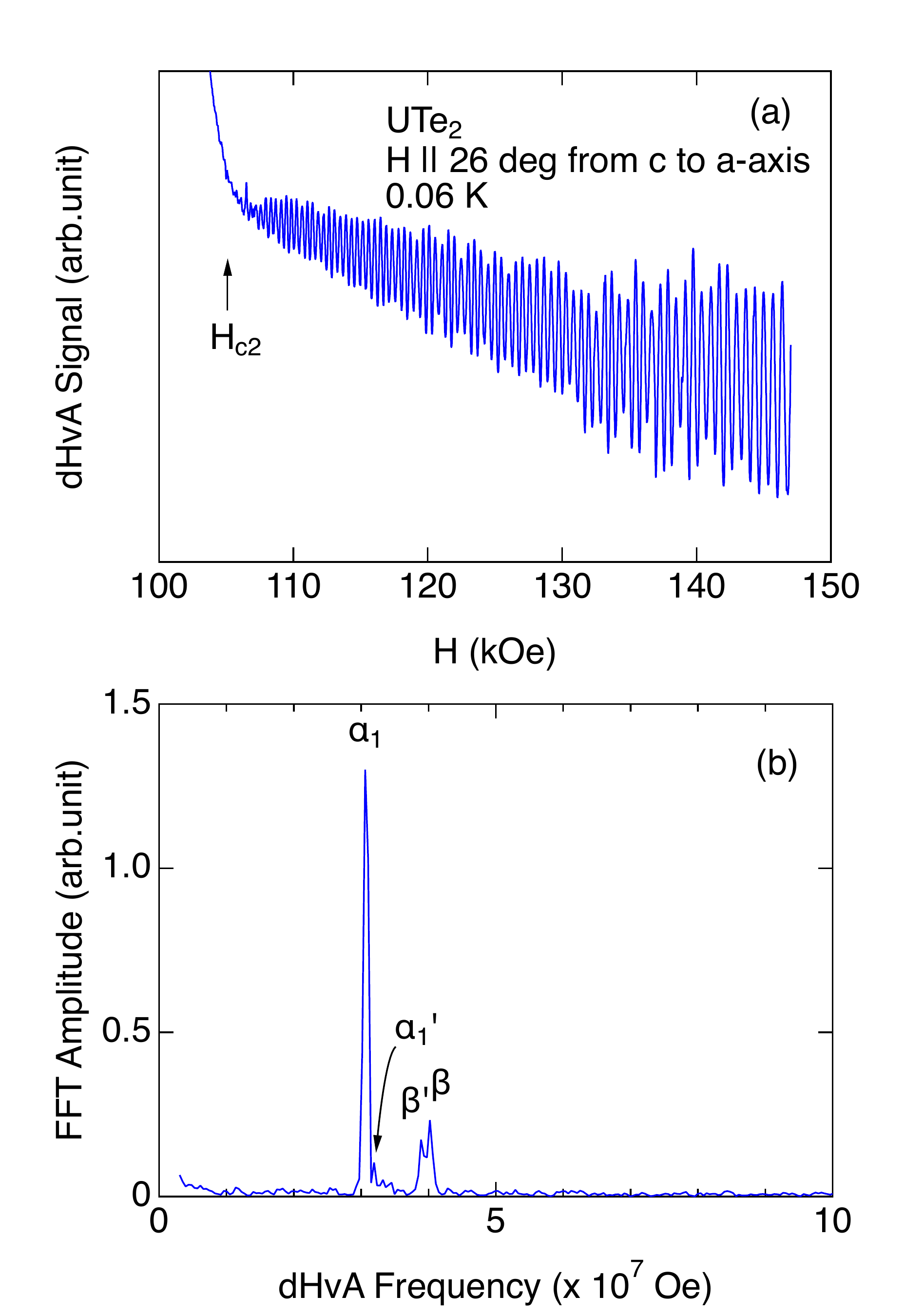}
\end{center}
\caption{(Color online) Typical dHvA oscillations and the FFT spectrum in the field range between $110$ and $147\,{\rm kOe}$ at the field direction tilted by 26 deg tilted from $c$ to $a$-axis in UTe$_2$ (sample\#1).}
\label{fig:UTe2_osc_FFT}
\end{figure}

Figure~\ref{fig:UTe2_osc_FFT} shows the typical dHvA oscillations and the corresponding FFT spectrum at the field angle of $26\,{\rm deg}$ tilted from $c$ to $a$-axis. 
Four dHvA frequencies, named $\alpha_1$, $\alpha_1^\prime$, $\beta$ and $\beta^\prime$ were detected. 

Figure~\ref{fig:UTe2_dHvA_AngDep}(a) shows the angular dependence of the dHvA frequencies from $c$ to $a$-axis. 
The results are obtained using two different samples, \#1 and \#2.
The sample \#1 was rotated from $H\parallel a$ to $c$-axis, while the sample \#2 was rotated from $H\parallel c$ to $a$-axis.
Both results are in good agreement with high reproducibility.
With increasing the field angle, branch $\alpha_1$ slightly decreases first, then increases up to the field angle $50\,{\rm deg}$.
Branch $\beta$ increases continuously up to $45\,{\rm deg}$.
Branch $\alpha_2$ also increases with the field angle, and shows a sharp increase around $60\,{\rm deg}$. 
The highest frequency reaches more than $1\times 10^8\,{\rm Oe}$, indicating a large cyclotron orbit.
No dHvA signal was detected around $H\parallel a$-axis, which is also confirmed in the field directions from $a$ to $b$-axis.~\cite{sup3} 
\begin{fullfigure}[tbh]
\begin{center}
\includegraphics[width= \hsize,clip]{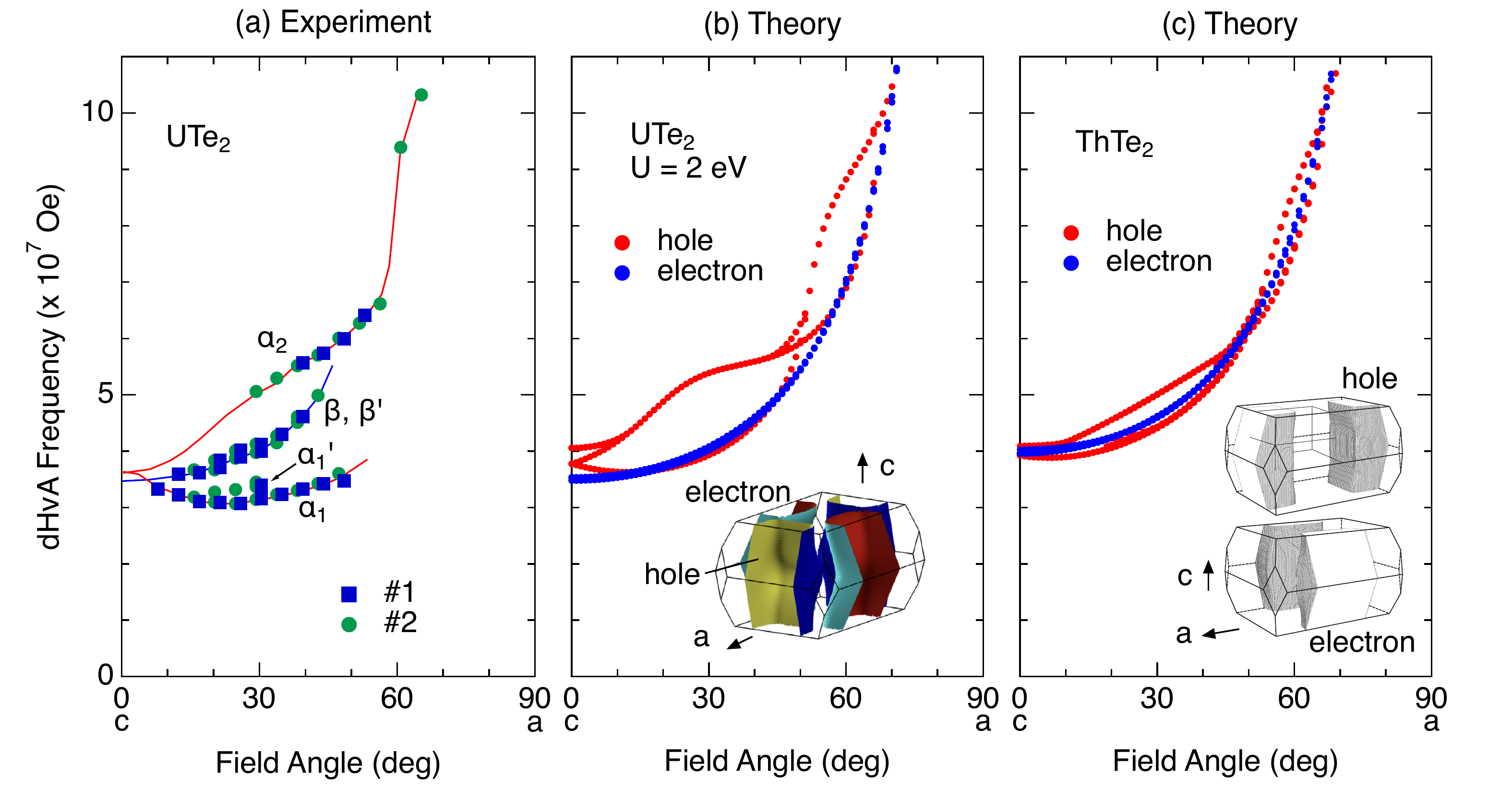}
\end{center}
\caption{(Color online) (a) Angular dependence of the dHvA frequency in UTe$_2$. Two samples \#1 (square), \#2 (circle) were used in different configurations. The lines are guides to the eyes. Panels (b) (c) show theoretical angular dependence of the dHvA frequency from the band calculations based on the GGA+$U$ ($U=2\,{\rm eV}$) method~\cite{Ish19} in UTe$_2$, and the LDA calculation in ThTe$_2$~\cite{Har20}, assuming the tetravalent U atom with the localized $5f^2$ configuration in UTe$_2$. The corresponding Fermi surfaces are depicted.}
\label{fig:UTe2_dHvA_AngDep}
\end{fullfigure}

In order to determine the cyclotron effective masses, the dHvA oscillations were measured at different temperatures.~\cite{sup4}
The results are summarized in Table~\ref{tab}.
The detected effective masses are very large in the range from $32$ to $57\,m_0$, 
indicating a direct evidence for a heavy electronic state from a microscopic point of view.

The Dingle temperature, $T_{\rm D}$, was also derived from the field dependence of the dHvA amplitude. 
At the field angle of $26\,{\rm deg}$, $T_{\rm D}$ for branch $\alpha_1$ is $0.16\,{\rm K}$.
From the simple relations, 
$F=\hbar c/(2\pi e) S_{\rm F}$,
$S_{\rm F}=\pi k_{\rm F}^2$,
$m_{\rm c}^\ast v_{\rm F} = \hbar k_{\rm F}$,
$T_{\rm D}= \hbar/(2\pi k_{\rm B}\tau)$,
and $l = v_{\rm F}\tau$,
where $S_{\rm F}$, $v_{\rm F}$, and $\tau$ are the cross-sectional area, Fermi velocity, and scattering life time, respectively,
we obtain the mean free path $l$ as $850\,{\rm \AA}$, indicating the high quality of our sample.

\begin{table}[h]
\newcommand{\N}{\phantom{0}}
\caption{Experimental dHvA frequency $F$, cyclotron effective mass $m_{\rm c}^\ast$, calculated dHvA frequency $F_{\rm b}$ and band mass $m_{\rm b}$ on the basis of the GGA+$U$ ($U=2\,{\rm eV}$) at the field angles tilted by $26$, $29$, $44$ and $61\,{\rm deg}$ from $c$ to $a$-axis in UTe$_2$.}
\label{tab}
\begin{center}
\begin{tabular}{ccccc}

\hline
				& \multicolumn{2}{c}{Experiment}				& \multicolumn{2}{c}{Theory}	\\
Branch 			& $F$ 					 & $m_{\rm c}^\ast$ 	&  $F_{\rm b}$			& $m_{\rm b}$\\ 
			    &($\times10^7\,{\rm Oe}$)& ($m_0$)			 	&($\times10^7\,{\rm Oe}$)& $(m_0)$    \\
\hline
$\theta=26\,{\rm deg}$\\
$\alpha_1$		& 3.08					& 32					& 3.80					& 3.0		\\
$\beta$			& 3.99					& 48					& 3.93					& 2.5		\\
\hline
$\theta=29\,{\rm deg}$\\
$\alpha_1$		& 3.15					& 33					& 3.89					& 3.1		\\
$\alpha_1^\prime$&3.38					& 40					& 						&    		\\
$\beta$			& 3.94					& 57					& 3.98					& 2.5		\\
$\beta^\prime$	& 4.09					& 34					& 4.05					& 2.6		\\
$\alpha_2$		& 5.05					& 39					& 5.36					& 2.4		\\
\hline
$\theta=44\,{\rm deg}$\\
$\alpha_1$		& 3.31					& 32					& 4.84					& 4.3		\\
$\alpha_2$		& 5.85					& 36					& 5.70					& 2.9    	\\
\hline
$\theta=61\,{\rm deg}$\\
$\alpha$		& 9.41					& 55					& 8.95					& 3.6		\\

\hline

\end{tabular}
\end{center}
\end{table}

The angular dependence of the dHvA frequencies are compared to those obtained from the calculations. 
Figures~\ref{fig:UTe2_dHvA_AngDep}(b) and \ref{fig:UTe2_dHvA_AngDep}(c) are the results from the GGA+$U$ ($U=2\,{\rm eV}$) calculation for UTe$_2$~\cite{Ish19} and from the LDA calculation for ThTe$_2$~\cite{Har20}, respectively.
The experimental results are fairly in good agreement with those of the GGA+$U$ ($U=2\,{\rm eV}$). 
Therefore we can assign the detected dHvA branches as follows.
Branch $\beta$ is ascribed to the electron Fermi surface with the cylindrical shape, giving rise to nearly the $1/\cos\theta$ dependence by tilting the field angle, $\theta$ from $c$ to $a$-axis.
Branches $\alpha_1$ and $\alpha_2$ originate from the same Fermi surface, that is a cylindrical hole Fermi surface.
Since the Fermi surface is corrugated from the cylinder shape, the dHvA frequency splits into $\alpha_1$ and $\alpha_2$,
which correspond to the minimal and maximal cross-sectional area, respectively, at low field angles,
when the field is titled from $c$ to $a$-axis.

The results of the LDA calculations in ThTe$_2$, which corresponds to U$^{4+}$ with the localized 5$f^2$ configuration in UTe$_2$,
shows a less agreement with the experimental results.
Nevertheless, two kinds of cylindrical Fermi surfaces are quite similar to those by GGA+$U$ ($U=2\,{\rm eV}$) as well as DFT with large $U$ ($U=7\,{\rm eV}$) calculations~\cite{Xu19}.
Note that the conventional LDA calculation predicts a Kondo semiconductor in UTe$_2$.~\cite{Aok19_UTe2,Har20,A.Shi19}
Similarly, GGA~\cite{Ish19} and DFT~\cite{Xu19} calculations without $U$ also shows a band gap at the Fermi energy.
These band structures are totally inconsistent with the experimental results, indicating that
the strong correlation should be taken into account in the calculations.
Small pocket Fermi surfaces predicted by other calculations~\cite{Fuj19,Shick21} are also inconsistent with our dHvA experiments.

Assuming the two kinds of cylindrical Fermi surfaces, which occupy approximately $20\,{\%}$ of volume for each in the Brillouin zone with the carrier compensation, one can roughly calculate the $\gamma$-value derived from each Fermi surface,
from the following equation, $\gamma = k_{\rm B}^2 V/(6\hbar^2) m_{\rm c}^\ast k_{\rm z}$.~\cite{Aok00c}
Here $V$ is the molar volume and $k_{\rm z}$ is the length of Brillouin zone along $c$-axis.
If we take $m_{\rm c}^\ast = 32\,m_0$ and $48\,m_0$ for the hole and electron Fermi surfaces, respectively,
the obtained $\gamma$-values are $40$ and $60\,{\rm mJ\,K^{-2}mol^{-1}}$ for each.
Thus, the total $\gamma$-value is $100\,{\rm mJ\,K^{-2}mol^{-1}}$, which
agrees with the value of $\gamma \sim 120\,{\rm mJ\,K^{-2}mol^{-1}}$ in the specific heat measurements,
indicating that our dHvA experiment detects the main Fermi surfaces of UTe$_2$.

The effective masses can be compared to the band masses from GGA+$U$ ($U=2\,{\rm eV}$) as shown in Table~\ref{tab}.
The band masses at the selected field angles are in the range from $2.5$ to $4.3\,m_0$, 
meaning that the mass enhancement, $m_{\rm c}^\ast/m_{\rm b}$ are approximately $10$--$20$.
This is also consistent with the mass enhancement, $\gamma / \gamma_{\rm b}$ obtained from the specific heat and the calculated density of states at the Fermi level ($\gamma_{\rm b}=8.1\,{\rm mJ\, K^{-2}mol^{-1}}$),
where the electron correlation is taken into account.

In the LDA calculations for ThTe$_2$ without the electron correlation, the band masses are much smaller.
For instance, $m_{\rm b}=0.7\,m_0$ is derived for the electron Fermi surface at $26\,{\rm deg}$,
while $m_{\rm c}^\ast = 48\,m_0$ is obtained in the dHvA experiments.
The mass enhancement, $m_{\rm c}^\ast/m_{\rm b}$ is consistent with that obtained from the Sommerfeld coefficient, namely $\gamma/ \gamma_{\rm b}$, in which $\gamma_{\rm b}$ is $1.7\,{\rm mJ\,K^{-2}mol^{-1}}$.

A question is whether the anisotropy of resistivity for $J\parallel a$, $b$ and $c$ can be explained by these Fermi surfaces,
because the resistivity for $J \parallel c$-axis, $\rho_c$ is only twice larger than that for $J\parallel a$-axis, $\rho_a$, and is comparable to that for $J\parallel b$-axis, $\rho_b$ at room temperature. ~\cite{Eo21}
At low temperature, the anisotropy increases, but not an order of magnitude.
In $\rho_c$, the rapid increase on cooling below $50\,{\rm K}$ with a maximum around $T^\ast\sim 15\,{\rm K}$ is observed,
suggesting that the electronic state may change from 3D to 2D-like nature on cooling. 
This is also in agreement with the development of the low dimensional antiferromagnetic fluctuation,
which only starts developing below $60\,{\rm K}$.~\cite{Kna21_PRB}

It should be noted that we cannot exclude the existence of small pocket Fermi surfaces with heavy masses,
which may induce a Lifshitz transition under magnetic field as proposed in thermopower experiments.~\cite{Niu20_PRL}
This may also compromise with possible topological superconductivity.

In summary, the dHvA oscillations were detected for the first time in UTe$_2$.
The angular dependence of the dHvA frequencies from $H\parallel c$ to $a$-axis,
are in good agreement with the results of GGA+$U$ ($U=2\,{\rm eV}$) based on the $5f^3$ itinerant model,
revealing two kinds of cylindrical Fermi surfaces from hole and electron bands.
The detected hole Fermi surface shows a large corrugation from the cylindrical shape.
On the other hand, the dHvA results are in less agreement with those of LDA calculations in ThTe$_2$, which
corresponds to U$^{4+}$ with the 5$f^2$-localized model.
The detected cyclotron effective masses are quite large, indicating heavy electronic states,
consistent with the $\gamma$-value of the specific heat.
These suggest the mixed valence states of U$^{4+}$ and U$^{3+}$, as proposed in the core-level spectroscopy~\cite{Fuj21},
which are sensitive to external parameters, such as pressure and field.
A link between our dHvA results on small energy scale near the Fermi level and the high energy spectroscopy~\cite{Fuj19,Fuj21,Mia20} deserves to be clarified.
Our results elucidate a key part to solve the UTe$_2$ properties,
as it happened when the Fermi surface was determined on high $T_{\rm c}$ superconductors~\cite{Doi07} and Ce-115 heavy fermion systems~\cite{Shi02}.

\section*{Acknowledgements}
We thank Y. \={O}nuki, S.-i. Fujimori, V. Mineev, Y. Tokunaga, M. Kimata, K. Ishida, K. Izawa, A. Miyake, J. P. Brison, D. Braithwaite, A. Pourret, I. Sheikin, and S. Fujimoto
for fruitful discussion.
This work was supported by KAKENHI (JP19H00646, JP20K20889, JP20H00130, JP20KK0061, JP22H04933), GIMRT (20H0406), ICC-IMR, and ANR (FRESCO).



\section*{Supplement}
\subsection*{Experimental}
The de Haas-van Alphen experiments were done using the field modulation technique at high magnetic fields up to $147\,{\rm kOe}$ and at low temperatures down to $70\,{\rm mK}$ in a top-loading dilution fridge.
The applied modulation field is up to $85\,{\rm Oe}$, and the signal was detected using a lock-in amplifier with $2\omega$-detection. 
The resistivity and specific heat was measured by the four-probe AC method down to $2\,{\rm K}$ and the thermal relaxation method down to $0.34\,{\rm K}$, respectively.
The AC susceptibility was measured using the same setup for the dHvA experiments with the small modulation field less than $10\,{\rm Oe}$.

\subsection*{Residual density of states and $T_{\rm c}$}
\begin{figure}[h]
\begin{center}
\includegraphics[width= 0.85\hsize,clip]{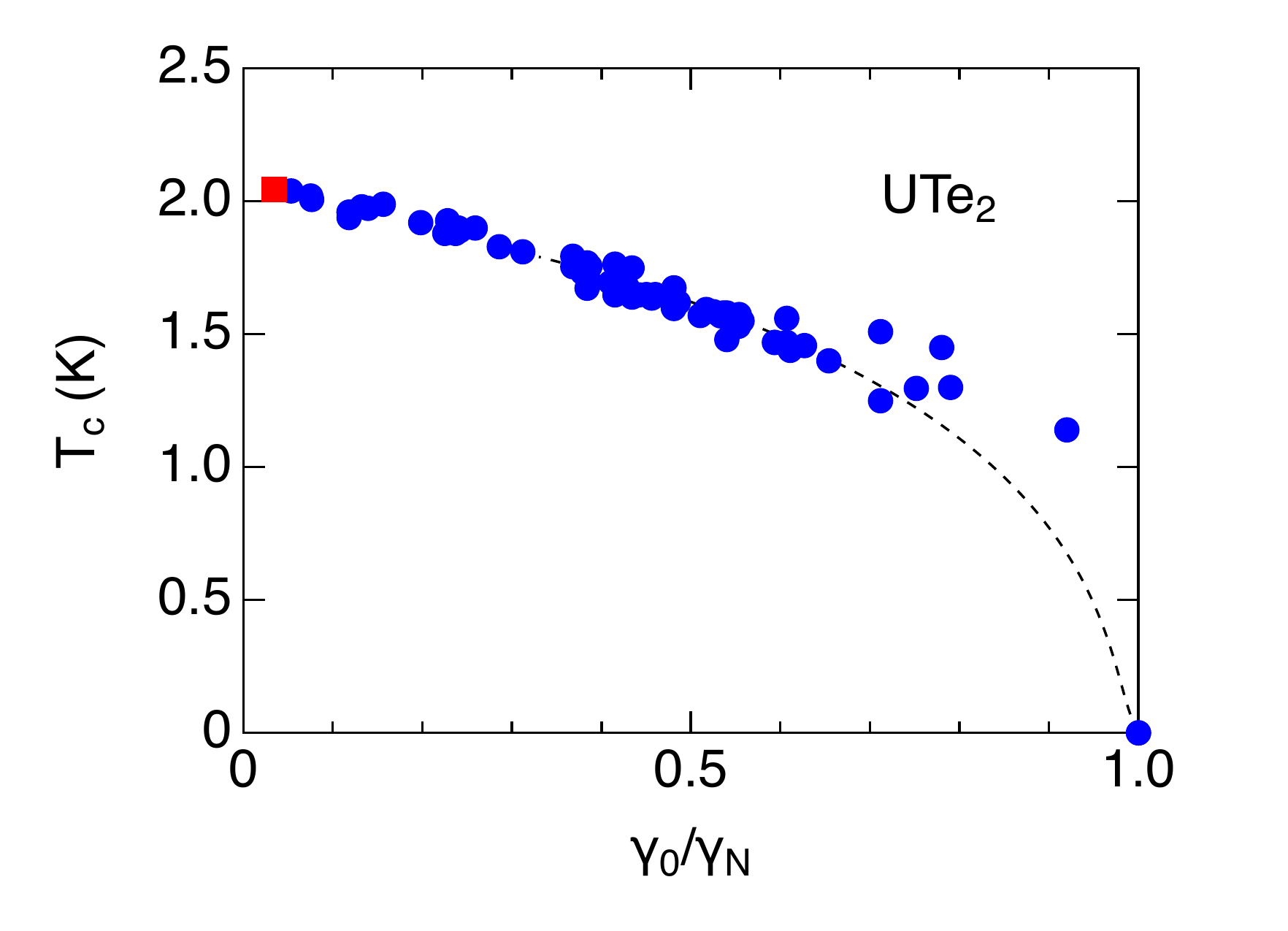}
\end{center}
\caption{(Color online) $T_{\rm c}$ vs residual $\gamma$-value ($\gamma_0$) scaled by that in the normal state ($\gamma_{\rm N}$) in different quality samples. The red square is sample\#1 measured in dHvA experiments, indicating the lowest $\gamma_0/\gamma_{\rm N} \sim 0.03$ and the highest $T_{\rm c}\sim 2.05\,{\rm K}$ among these different samples.}
\label{fig:UTe2_suppl_gamma0}
\end{figure}

Figure~\ref{fig:UTe2_suppl_gamma0} shows $T_{\rm c}$ as a function of the scaled $\gamma$-value, $\gamma_0/\gamma_{\rm N}$ for different quality samples, determined by the specific heat measurements. 
Our sample is located at the lowest $\gamma_0/\gamma_{\rm N}$ and the highest $T_{\rm c}$,
indicating the highest quality among these different quality samples.~\cite{Aok22_UTe2_review}

\subsection*{FFT spectra at different temperatures and mass plot}
\begin{figure}[tbh]
\begin{center}
\includegraphics[width= 0.8\hsize,clip]{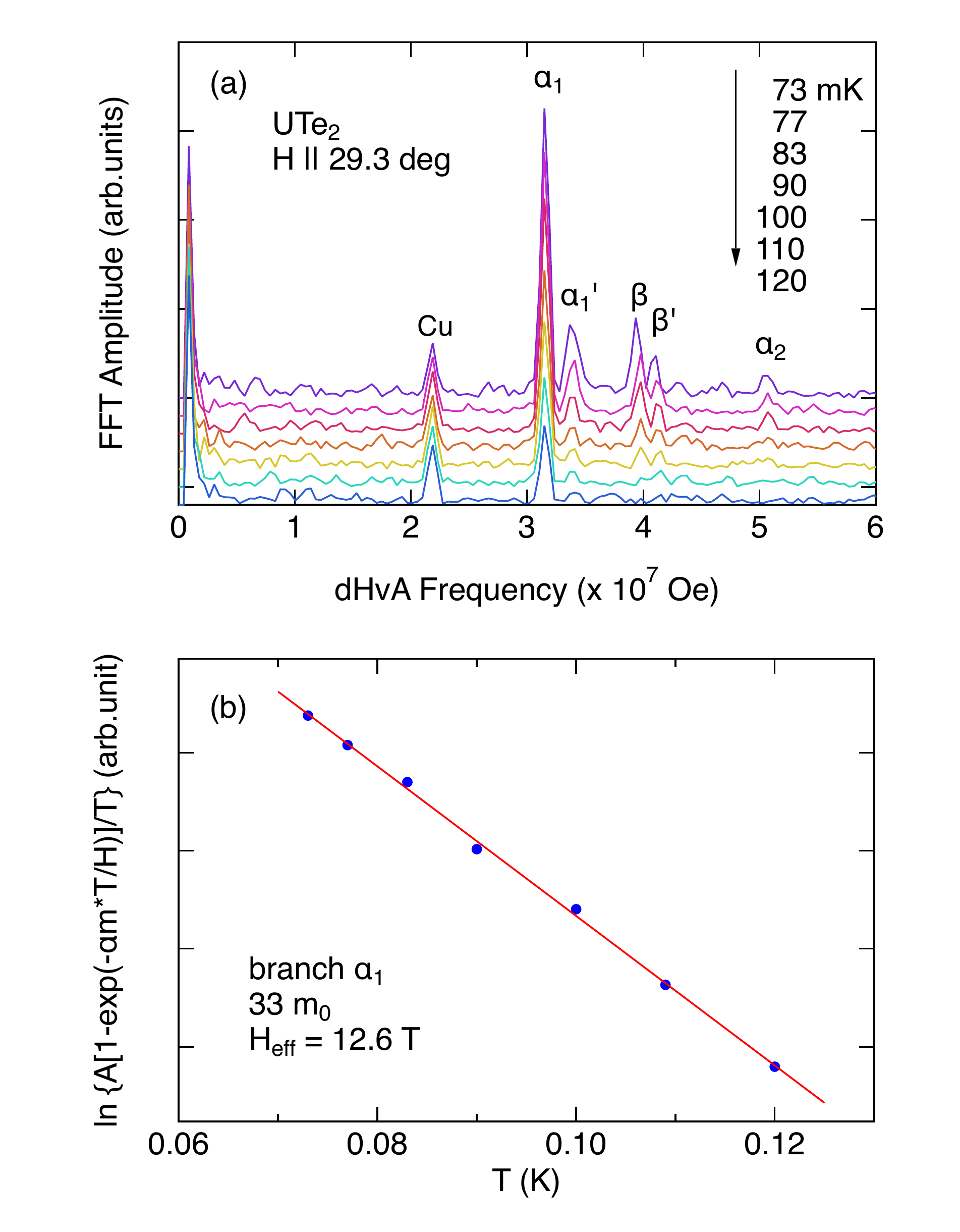}
\end{center}
\caption{(a) FFT spectra at different temperatures up to $120\,{\rm mK}$ for the field tilted by $29.3$ deg from $c$ to $a$-axis in UTe$_2$ (sample\#2). The data are shifted upward for clarity.
Five fundamental dHvA frequencies originating from two kinds of cylindrical Fermi surfaces are detected. The frequency, $2.1\times 10^7\,{\rm Oe}$ corresponds to the neck orbit of Copper from the pick-up coil. (b) The mass plot for branch $\alpha_1$.}
\label{fig:UTe2_mass_plot}
\end{figure}

Figure~\ref{fig:UTe2_mass_plot}(a) shows the FFT spectra at a field angle of $29\,{\rm deg}$ for different temperatures. 
The FFT amplitudes for branch $\alpha_1$, $\alpha_1^\prime$, $\beta$, $\beta^\prime$, and $\alpha_2$ are strongly suppressed already at $100\,{\rm mK}$,
suggesting the heavy effective masses, while the amplitude of $F=2.1\times 10^7\,{\rm Oe}$ originating from the neck orbit of Copper from the pick-up coil is almost unchanged because of the light effective mass.
The mass plot for branch $\alpha_1$ is shown in Fig.~\ref{fig:UTe2_mass_plot}(b), which derives the large effective mass of $33\,m_0$.

\subsection*{Angular dependence of dHvA frequencies: comparison to GGA+$U$ at different $U$}
For the field directions close to the $a$-axis, no dHvA oscillations are detected. 
This is also confirmed by the field rotation from $a$ to $b$-axis.
In the GGA+$U$ with the smaller $U=1.1$ and $1.5\,{\rm eV}$, the dHvA frequency ($\sim 2.5\times 10^7\,{\rm Oe}$) is predicted for the field direction close to the $a$-axis, as shown in Figs.~\ref{fig:UTe2_suppl_AngDep}(c)(d). 
This dHvA branch originates from the connected electron Fermi surface at the $X$ point, forming a ring-shaped Fermi surface instead of the disconnected cylindrical Fermi surface. 
The band mass is about $3\,m_0$ and the curvature factor is preferable for detecting the dHvA oscillations.
However, no dHvA oscillations were experimentally detected.
Therefore it is expected that main Fermi surfaces consist of the two kinds of cylindrical Fermi surfaces.
\begin{fullfigure}[tbh]
\begin{center}
\includegraphics[width=\hsize,clip]{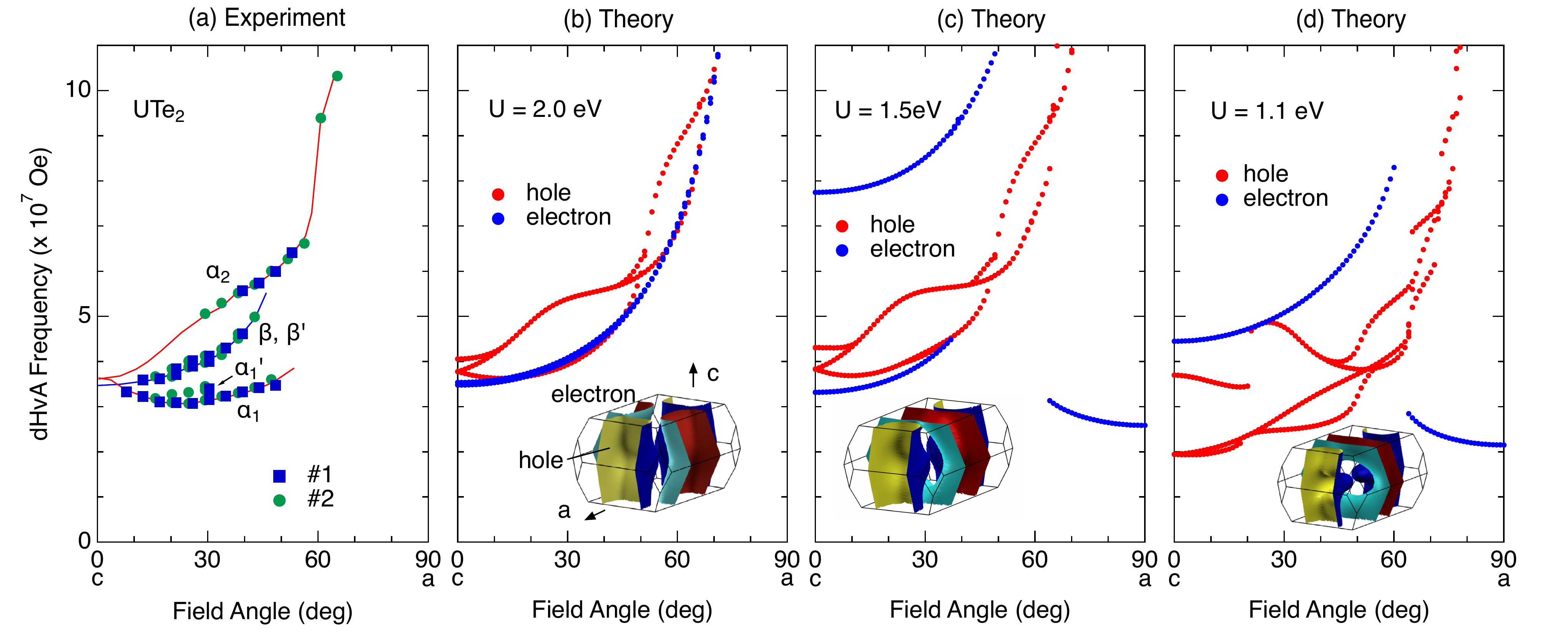}
\end{center}
\caption{(a) Angular dependence of the dHvA frequencies in UTe$_2$. (b)-(d) Theoretical angular dependence of the dHvA frequencies calculated by the GGA+$U$ methods with different $U$, $U=2.0$, $1.5$ and $1.1\,{\rm eV}$. The corresponding Fermi surfaces are depicted.}
\label{fig:UTe2_suppl_AngDep}
\end{fullfigure}

\end{document}